%% file: main-template.tex
\documentclass[a4paper]{article}

\usepackage{INTERSPEECH2022}

\usepackage{booktabs} 
\usepackage{multirow}
\usepackage{algorithm}
\usepackage{algorithmic}
\usepackage[font=small]{caption}
\usepackage{xcolor}

\title{Latency Control for Keyword Spotting}
\name{Christin Jose, Joseph Wang, Grant P. Strimel, Mohammad Omar Khursheed, Yuriy Mishchenko, Brian Kulis}
\address{
  Amazon Science, United States}
\email{\{chrjse,wangjose,gsstrime,khursmoh,yuriym,kulibria\}@amazon.com}

\begin{document}









\maketitle
\begin{abstract}
Conversational agents commonly utilize keyword spotting (KWS) to initiate voice interaction with the user. For user experience and privacy considerations, existing approaches to KWS largely focus on accuracy, which can often come at the expense of introduced latency. To address this tradeoff, we propose a novel approach to control KWS model latency and which generalizes to any loss function without explicit knowledge of the keyword endpoint. Through a single, tunable hyperparameter, our approach enables one to balance detection latency and accuracy for the targeted application.
Empirically, we show that our approach gives superior performance under latency constraints when compared to existing methods.
Namely, we make a substantial 25\% relative false accepts improvement for a fixed latency target when compared to the baseline state-of-the-art.
We also show that when our approach is used in conjunction with a max-pooling loss, we are able to improve relative false accepts by 25\% at a fixed latency when compared to cross entropy loss. 

\end{abstract}

\maketitle

\input{main-content}

\bibliographystyle{IEEEtran}


\bibliography{main-bibliography} 

\end{document}

%% file: main-content.tex

\section{Introduction}

Keyword spotting (KWS) is the task of detecting keywords of interest in a continuous audio stream. One common application of keyword spotting is as a means to interact with a voice assistant such as Amazon Alexa, Google Assistant, or Apple Siri, where saying the keyword initiates a dialogue. In this application, the KWS model acts as a gating mechanism, where detection of the keyword triggers more computationally intensive processing of the audio (occurring either on the device or with audio streamed to a cloud-side system) to extract meaning and elicit a response from the assistant \cite{Macoskey2021BifocalNA}.  Such processing requires the KWS model to be both accurate and have low latency, that is, to detect the keyword as soon as possible after it is spoken. This dual objective spurs a natural trade-off between accuracy and latency.  Low accuracy can lead to a failure to respond or unprompted actions or responses, while increased latency not only causes delays in responses but can change user behavior, such as causing a user to repeat the keyword while waiting for a response.

One classic approach to keyword detection is through a 2-stage model comprised of a first-stage Deep Neural Network (DNN) acoustic model and a Hidden Markov Model (HMM) \cite{Panchapagesan2016MultiTaskLA, Compressedinproceedings, Sun2017AnES, Kumatani2017DirectMO, Bottleneckinproceedings}. Such a keyword spotter may also have additional classifiers after the HMM, such as support vector machines, to increase the accuracy of keyword detection \cite{minhua1}. These KWS systems naturally provide the endpoints of the keyword via the HMM output, and due to their training on subword components, they generally have low latency. During runtime, these systems perform Viterbi decoding of the keyword senone sequences that produces the times of the keyword's start and end senones in the input audio. However, this procedure can be computationally expensive, potentially require subword labels (e.g. labeling of individual phones), and can be difficult to train and tune efficiently.

As a result, more recent work has focused on building fully neural-based detection models that omit HMM's from the detection process. A common approach is to employ a single-stage feed-forward DNN \cite{Smallfootprint1} where the model is trained to directly predict the keyword (or individual words in a phrase keyword). This approach is attractive for running on hardware-limited devices as the memory and CPU requirements can easily be tuned, can be trained with only labeling of keywords, and has been shown to outperform a keyword/filler HMM approach. Similarly, Convolutional Neural Networks (CNNs) \cite{sainath15} and Convolutional Recurrent Neural Networks (CRNNs) \cite{omar_asru, Khursheed2020SmallFC} have shown improvements over fully connected feed-forward DNN's with similar advantages related to ease of training and compute tuning.

The max-pooling loss \cite{Sun2016MaxpoolingLT} is a recently proposed approach to training neural KWS models. This loss selects a frame automatically to minimize the cross-entropy loss.  For a positive example, it selects the frame that has the maximum score on the corresponding class, and for a negative example, it selects the frame that has the lowest score. By constructing the loss with a max-pooling, the model is trained without the need for labeling either subword components or the exact alignment of the keyword. Empirically, this has been shown to outperform models trained on pre-selected examples, where the alignment of the keyword segment is determined during dataset creation. The method that used to select the keyword alignments directly influences the KWS model performance and thus not needing precise alignment information helps in improving detection accuracy.

However, one issue that arises from the max-pooling loss is an increase in detection latency. We hypothesize that the model can leverage the additional audio context to improve detection accuracy and thereby increasing the detection latency. Increased latency is particularly noticeable in the case when the audio context for a single decision for the KWS model is significantly larger than the spoken keyword, such as when a user has a high speaking rate. In this case, the max-pooling loss introduces no incentive to detect the keyword immediately following the utterance of the keyword, and as a result, the model instead often outputs the maximum after receiving additional audio context following the keyword. While this additional audio context can help improve model performance, the introduced latency can significantly degrade the customer experience and disrupt the interaction between the user and a conversational agent.  Although the increased latency associated with the max-pooling loss could be addressed by introducing a penalty on model latency, this requires accurate labeling of the keyword end-point, removing one of the key advantages of the max-pooling loss.

In this paper we introduce an approach to training KWS models using a loss function that allows for latency control without labeling of the exact alignment of the keyword. The proposed loss function is a generalization of the max-pooling loss that shifts the alignment selected for positive examples to an earlier frame based on a probability, forcing the model to emit earlier detection peaks. By changing this probability, the balance between KWS model accuracy and latency can easily be adjusted without accurate knowledge of the keyword endpoint.

Empirically, we show this approach yields models outperforming the state-of-the-art (SOTA) models for varying latency constraints despite not relying on accurate labeling of the keyword endpoint required for the existing SOTA approach. By adjusting a single parameter of the loss, we are able to meet various latency constraints, with the proposed approach outperforming the SOTA approach for all latency-constrained scenarios. Furthermore, the monotonic relationship between the loss parameter and latency allows for latency constraints to be easily met with limited exploration of the parameter space and thus limited additional training time.

\section{Related Work}\label{sec:relatedWork}

Most of the literature on KWS has historically focused largely on accuracy improvement and novel architectures \cite{kws_1, kws_2, kws_3, kws_4}, though more recent work has also explored latency control. For instance, the accuracy-latency trade-off was discussed by Sigtia et al. \cite{Sigtia}. Here, a second stage verification model that uses more context after the keyword to detect hard examples was shown to improve model accuracy at the cost of additional latency for some examples. This approach is in contrast to our motivation here where we aim to train a single stage model providing the capability to the user to choose an operating point for accuracy and latency. From a neural compute reduction perspective, architectural proposals, such as low-latency CRNN \cite{lcrnn}, have been applied for reductions in parameter count and inference operations, which ultimately translate into latency improvements on compute constrained hardware. 
Here, however, we focus on the \textit{algorithmic latency} for KWS, measured by the number of frames after the keyword endpoint the model emits a decision, rather than overhead latency introduced by compute limitations. Notably, Fu et al. \cite{Gengshen} propose a loss function to control detection latency by masking audio context after the keyword during training. While sharing similarities with our method, this approach requires explicit knowledge of the keyword endpoint for training, however, and does not easily generalize to other loss functions. These limitations are the central motivators for our method.  We use the loss of \cite{Gengshen} as the SOTA baseline to compare our results within this paper and refer to their method as the ``max-latency approach''.

\begin{figure}[!htb]
\centering
\includegraphics[width=0.40\textwidth]{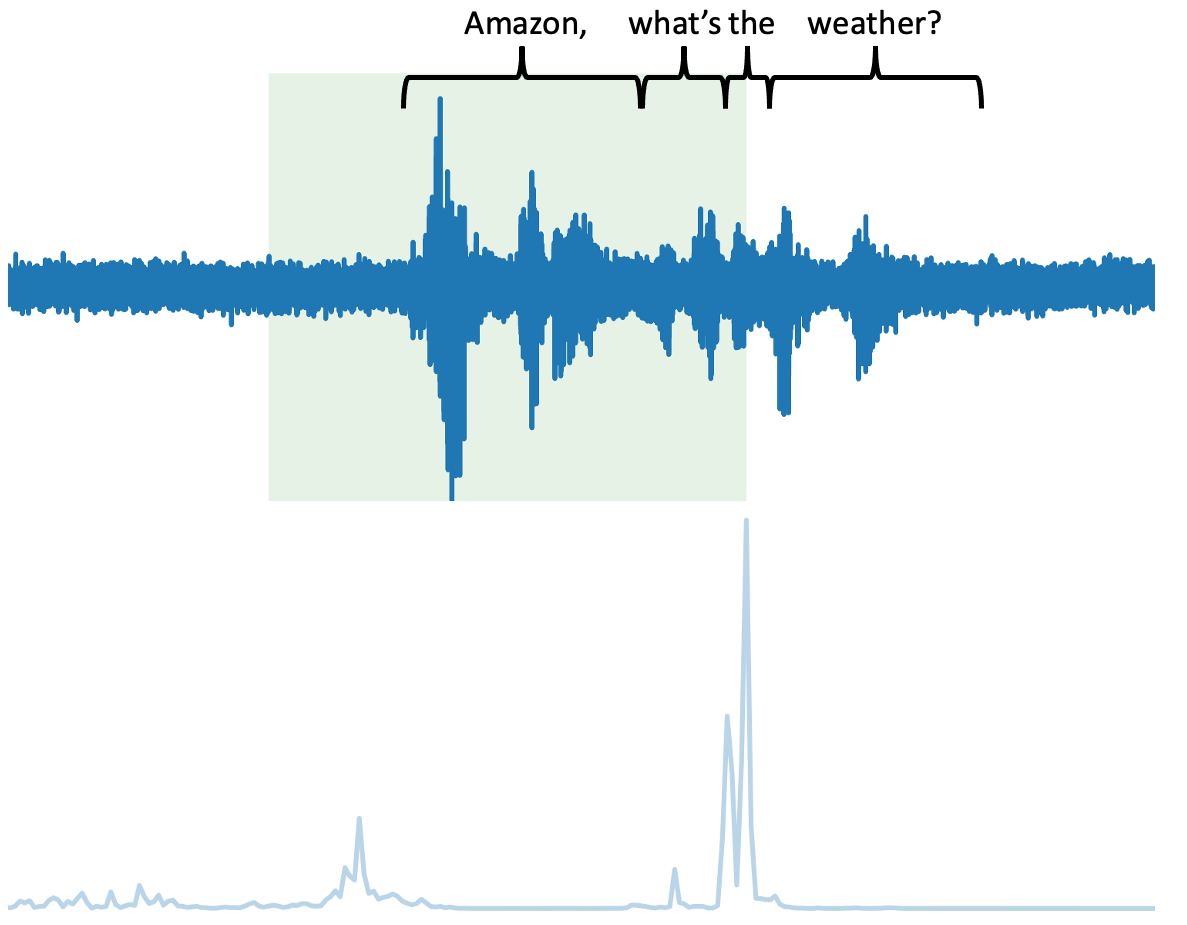}
\caption{Audio of the author saying ``Amazon, what's the weather'' is shown, the aligned posteriors from a model trained using the max-pooling loss shown below. The context window for the peak model posterior is overlaid on the audio signal in green.}
\label{fig:max_pool_example}
\vspace{-4mm}
\end{figure}

\begin{figure*}[!htbp]
\centering
\includegraphics[width=1\textwidth]{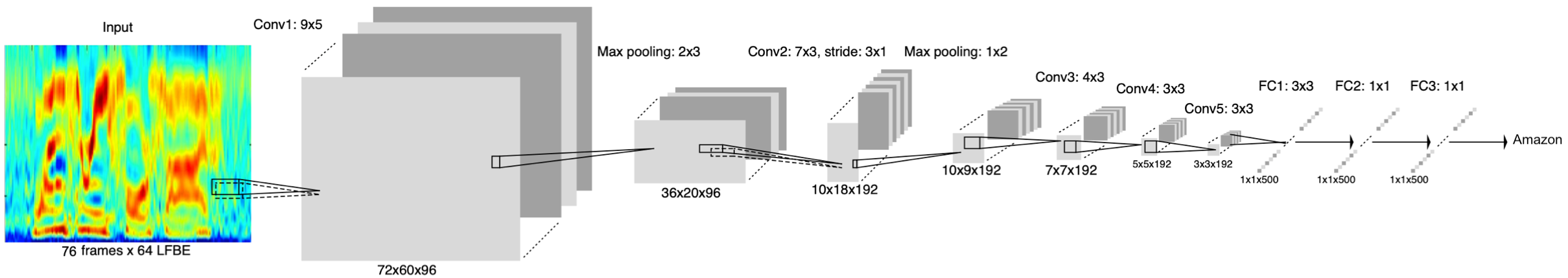}
\caption{CNN architecture used in the experiments. Input window consists of 76 frames of audio.}
\label{fig:cnn_diagram.png}
\vspace{-3mm}
\end{figure*}

\section{Latency Control for Keyword Spotting}\label{sec:wordLevel}
In this section we give a brief introduction on max-pooling loss after which we dive into the proposed approach to control keyword detection latency.

\subsection{Max Pooling Loss}\label{sec:maxPooling}

The max-pooling loss, given in Eqn. \ref{maxpooling_ce_loss} based on cross-entropy, was proposed in \cite{Sun2016MaxpoolingLT} for training RNN KWS models:
\begin{align} \label{maxpooling_ce_loss}
\mathcal{L}&(y, [p_{y0},\ldots,p_{yT}]) = -\log(p_{yt}), \\ 
t &= \begin{cases}
\underset{i \in \{0, \ldots, T\}}{\text{argmin}} \: (p_{0i}) & y = 0\\
\underset{i \in \{0, \ldots, T\}}{\text{argmax}} \: (p_{yi}) & y \neq 0
\end{cases} \nonumber
\end{align}
where $y$ is the example label and $p_{yt}$ is the posterior for class $y$ output by the model at time $t$.

In contrast to traditional classification based on minimizing cross-entropy loss for a specific alignment (selected as part of the data preparation process), the max-pooling loss selects a
frame automatically on which it minimizes the cross-entropy
loss. For a positive example, it selects the frame that has the
maximum score on the corresponding class, since we only
need the highest score of the class to pass a threshold to detect the keyword. For a
negative example, it selects the frame that has the lowest score
on the negative class (equivalent to the highest score on the positive class in the binary classification setting), encouraging low positive scores across all frames by maximizing
the lower bound of negative scores.

Allowing the model to choose the maximal alignment presents two main advantages over using data where the keyword alignment is manually selected. First, the alignment information necessary on training data is significantly reduced, as only a noisy estimate of the location needs to be used to crop a segment of audio containing (but not exactly aligned with) the keyword. This can significantly reduce both data preparation/annotation overhead as well as avoid errors introduced by poor alignment such as clipping of the keyword. Second, allowing the model to select the maximum posterior leads to improved empirical performance as compared to pre-selected alignments \cite{Sun2016MaxpoolingLT}.

Unfortunately, by allowing the model to determine the maximum frame in the case of a positive example, a significant amount of latency can be introduced into the model, especially when the input segment has a non-trivial amount of audio following the keyword, where we hypothesize the model can leverage this additional audio context to improve detection accuracy. An illustration of this behavior can be seen in Fig. \ref{fig:max_pool_example}. Here, the model correctly detects the keyword, however, does so with significant latency. Intuitively, this is unsurprising as the model likely is able to generate a more confident prediction given the audio context following the keyword.

\subsection{Latency Control}\label{sec:latencyControl}

One possible approach to avoid this introduced latency is to clip the training audio segment immediately following the keyword; however, this requires knowledge of the endpoint of the keyword. In the extreme case, limiting the audio following the keyword endpoint reduces to the cross-entropy loss where the model is trained on manually aligned data.

Rather than limiting the context of the audio data, we propose a simple augmentation of the max-pooling loss in order to reduce detection latency:
\begin{align}\label{latency_ce_maxpooling_loss}
\mathcal{L}&(y, [p_{y0},\ldots,p_{yT}]) = -\log(p_{yt}), \\ 
t &= \begin{cases}
\underset{i \in \{0, \ldots, T\}}{\text{argmin}} \: (p_{0i}) & y = 0\\
\left(\left(\underset{i \in \{0, \ldots, T\}}{\text{argmax}} \: (p_{yi})\right) - \beta\right)_{+} & y \neq 0
\end{cases} \nonumber
\end{align}

where the function $(\cdot)_+$ is defined $(x)_{+} = \max (x, 0)$ and $\beta$ is a random value drawn from a predefined distribution $\mathcal{B}$. In practice, we set $\mathcal{B}$ to a Bernoulli distribution with user-defined parameter $b$ (the probability that $\beta=1$).

For positive examples, with the probability $b$, the loss is computed not on the maximum class posterior frame but instead on the posteriors generated by the prior frame. Intuitively, by randomly evaluating the loss of the model on an earlier posterior, we expect that the posterior prior to the peak detection will also be high, allowing for earlier detection. As the user increases the parameter $b$, an increasingly larger emphasis is placed on the prior posterior. Through varying this parameter, the user can balance between detection accuracy and detection latency.

Note that in this application, the model posterior stride is relatively large, so pushing the model to use the prior posterior through use of a Bernoulli distributed value $\beta$ is sufficient to reduce latency; however, for detection models with smaller times between posteriors or where larger latency reductions are needed, positive distributions such as a Poisson distribution could be used.

Furthermore, we note that the loss in Eqn. \ref{latency_ce_maxpooling_loss} can be generalized to any loss function that takes in the posteriors from a single time point and label using the same approach to selection of alignment:
\begin{align}\label{latency_gen_maxpooling_loss}
\mathcal{L}&(y, [P_{0},\ldots,P_{T}]) = L\left(y, P_{t}\right), \\ 
t &= \begin{cases}
\underset{i \in \{0, \ldots, T\}}{\text{argmin}} \: (p_{0i}) & y = 0\\
\left(\left(\underset{i \in \{0, \ldots, T\}}{\text{argmax}} \: (p_{yi})\right) - \beta\right)_{+} & y \neq 0
\end{cases} \nonumber
\end{align}
where $P_{i}$ is the set of posteriors across class labels at time $i$ and $L(y, P)$ is a loss function between a true label $y$ and vector of model posteriors $P$. Empirically, we find cross-entropy to be highly effective for the KWS task; however, given differing model requirements (e.g. varying importance of individual classes or differing costs per error type), the proposed approach can easily be generalized to a wide class of problems.

One primary advantage of the proposed approach over prior work is no estimate of the keyword endpoint is necessary. This approach allows for the same dataset, training procedure, and model architecture as used for the max-pooling loss, with only a simple, computationally efficient modification of the loss function applied.

\section{Experiments}

In this section, we evaluate the method that we propose against the baseline methods. The metric that we use of latency is detection time relative to the end index of the keyword in milliseconds (ms). Negative latency shows that the keyword was detected before its completion. For studying accuracy of detection, we use relative false accept and relative false reject counts to plot the detection error tradeoff (DET) \cite{det} curves for the keyword in the analysis. All models were trained on the same corpus of de-identified, annotated data composed of approximately 20,000 hours of training data, a fraction of which was used for hyperparameter tuning. Individual keyword example segments were roughly aligned using an HMM keyword model which naturally provides an estimate of the alignment for a keyword. Additional padding before and after the keyword was added to ensure alignment errors did not lead to clipping.

\begin{figure}[!htb]
\centering
\includegraphics[width=0.46\textwidth]{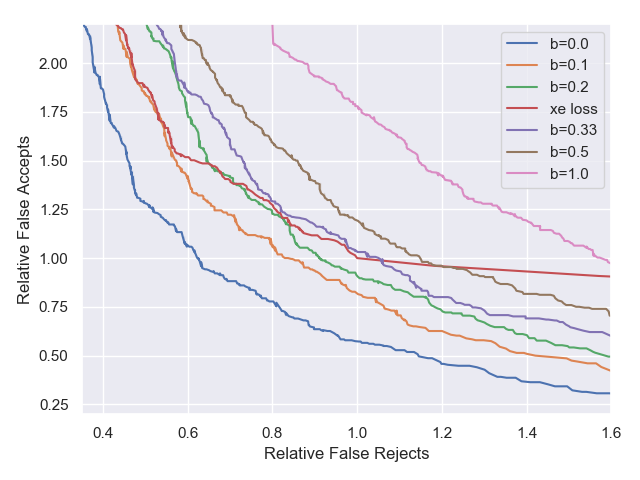}
\caption{DET curves for models trained using the proposed approach with various parameter values $b$ as well as a model trained using HMM-aligned data (labeled ``xe loss.''. Relative performance is reported with respect to the ``xe loss'' model.}
\label{fig:combined_det_curves}
\vspace{-3mm}
\end{figure}

\begin{figure}[!htb]
\centering
\includegraphics[width=0.46\textwidth]{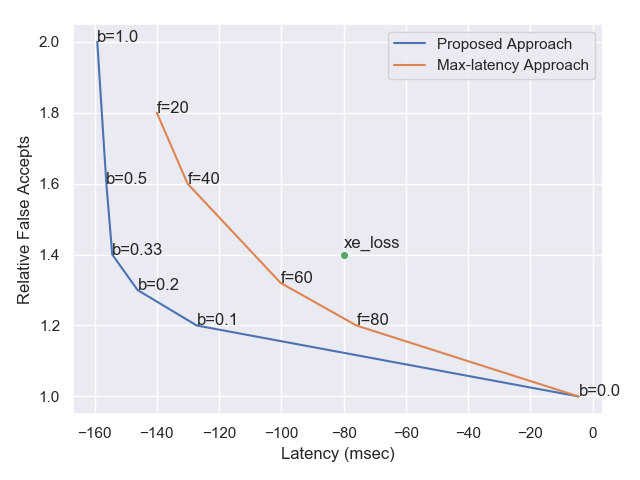}
\caption{Latency vs. relative false accepts for the proposed approach (in blue), the max-latency approach \cite{Gengshen}, and the model trained using HMM-aligned data (labeled ``xe loss.'') Performance is given with respect to the max-pooling loss ($b=0.0$), with the latency reported as the reduction relative to this model.}
\label{fig:compare_with_gengshen.png}
\vspace{-3mm}
\end{figure}

\begin{figure}[!htb]
\centering
\includegraphics[width=0.49\textwidth]{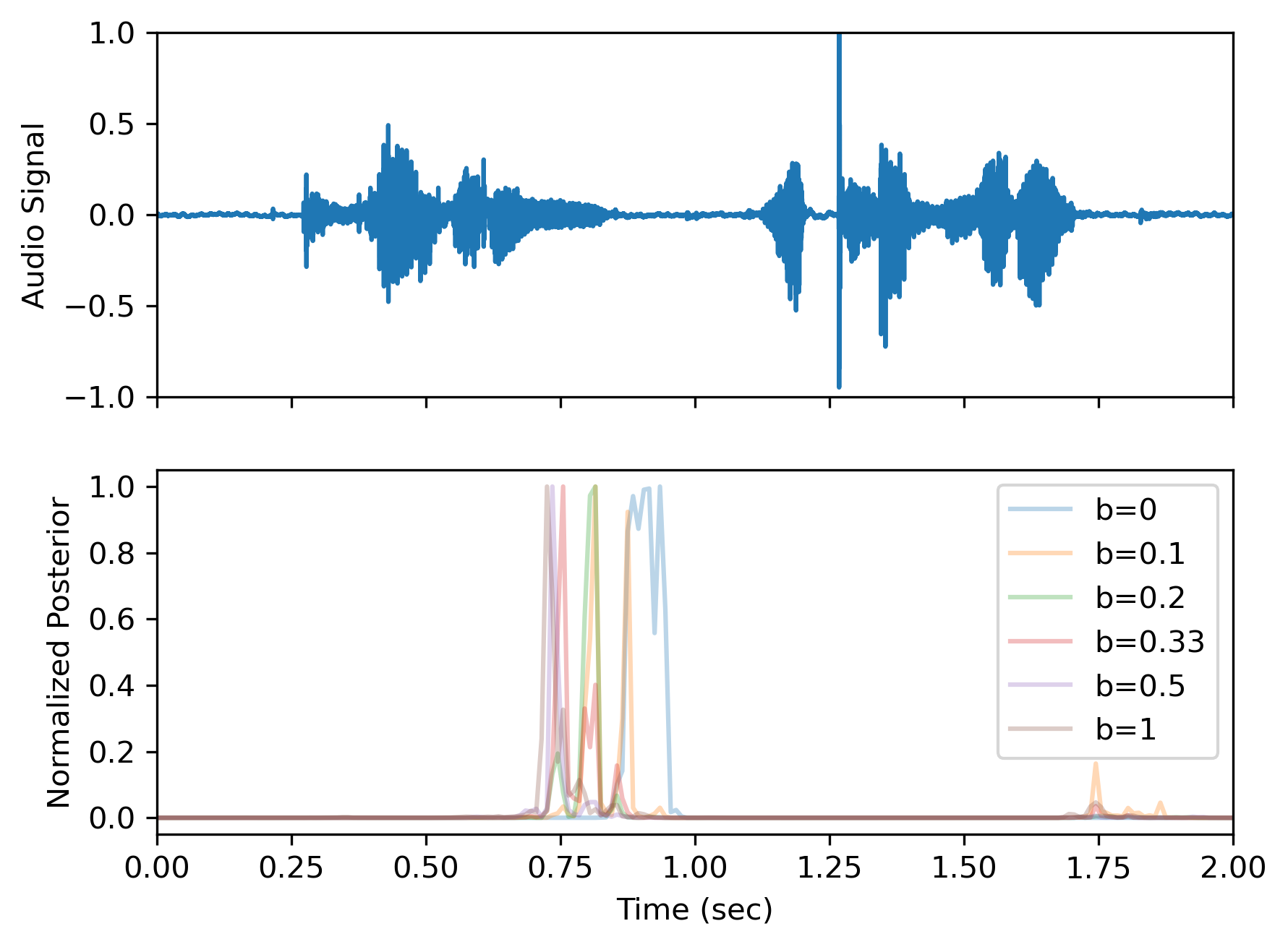}
\caption{Audio signal and aligned posteriors for models trained with different values of $b$.}
\label{fig:all_score_plots.png}
\end{figure}

We train a variety of models using the CNN architecture proposed in \cite{jose20_interspeech} (Fig. \ref{fig:cnn_diagram.png}) for the binary keyword detection problem for the keyword \textit{Amazon}. Audio is first converted into 64-dimensional log Mel-filterbank energy (LFBE) features calculated over 25ms frames with a 10ms shift before being passed to the model. The KWS models are composed of five convolutional layers followed by three fully connected (FC) layers with max-pooling after the first layer and 3-stride convolution in the second layer. We train model using Adam optimizer and a dropout rate of 0.3 for all the hidden layers.

The same initialization and training approach was used for all models, with only the loss functions varying across models. Evaluation was performed over approximately 1,400 hours of annotated, de-identified data.

We trained 6 models using the proposed approach with parameter values of $b \in \{0.0, 0.1, 0.2, 0.33, 0.5, 1.0\}$, where $b=0$ is equivalent to the standard max-pooling loss.  Additionally, we train a model using a cross-entropy loss without max-pooling, where the alignment for each utterance of the keyword is based on the outputs of the HMM model. The DET curves for these models are shown in Fig. \ref{fig:combined_det_curves}, where the relative rate of false accepts and false rejects compared to the baseline cross-entropy trained model are shown.

As expected, the max-pooling model ($b=0$) has the best detection performance, with the performance decreasing as the parameter $b$ increases, matching the cross-entropy detection performance when $b=0.33$ and for larger values of $b$ there is further degradation in detection performance as the importance of latency increases at the expense of detection performance.

Additionally, we train models using the max-latency approach proposed in \cite{Gengshen} sweeping across the maximum latency parameter, $f$, used to control detection latency. This parameter limits the audio context used in the model following the keyword endpoint as determined by the HMM model. This is equivalent to reducing the range of posteriors included in the argmax in Eqn. \ref{maxpooling_ce_loss}, forcing the posterior peak to occur earlier to not incur a large loss.

To demonstrate the trade-off between performance and latency, we plot the relative latency of the models compared to the relative false accept rates of the models at a fixed false reject rate in Fig. \ref{fig:compare_with_gengshen.png}, with the max-pooling loss used as the baseline due to having the highest latency and lowest false accept rate. Note that the latency shown along the x-axis is the average reduction in detection relative to the max-pooling loss at the thresholds selected for the fixed false rejection rate. Our proposed approach clearly outperforms both the max-latency approach as well as the model trained using a cross-entropy loss on predetermined alignments and additionally has the benefit of not requiring exact estimates of the keyword alignment during training.

We also provide a visualization showing the impact of varying the parameter $b$ on a single example of audio in Fig. \ref{fig:all_score_plots.png}. Here, a clean audio segment of the user saying \textit{``Amazon, what time is it?''} is plotted, with the normalized detection posteriors plotted below for models trained with the proposed approach. As the value of $b$ is increased, the posterior peak is shifted earlier and earlier, until in the extreme case of $b=1$, the posterior peak occurs prior to the completion of the keyword, demonstrating the ability to reduce latency simply by varying the parameter $b$.

\section{Conclusion}

Latency is an important aspect of KWS models for use in conversational agent systems. In this paper we present an approach to reduce KWS latency through use of a novel loss function. Our proposed approach does not require knowledge of the keyword endpoint, is easily implementable, and yields models outperforming SOTA approaches that rely on knowledge of the keyword endpoint. Our results show that we make a significant 25\% relative false accepts improvement over the baseline model at fixed latency. As a future work we plan to experiment with this hyperparameter in other commonly used loss functions like the cross entropy loss, and extends its applicability to other use cases such as utterance detection in signal-to-interpretation applications.


%% file: main-template.bbl
\begin{thebibliography}{10}
\providecommand{\url}[1]{#1}
\csname url@samestyle\endcsname
\providecommand{\newblock}{\relax}
\providecommand{\bibinfo}[2]{#2}
\providecommand{\BIBentrySTDinterwordspacing}{\spaceskip=0pt\relax}
\providecommand{\BIBentryALTinterwordstretchfactor}{4}
\providecommand{\BIBentryALTinterwordspacing}{\spaceskip=\fontdimen2\font plus
\BIBentryALTinterwordstretchfactor\fontdimen3\font minus
  \fontdimen4\font\relax}
\providecommand{\BIBforeignlanguage}[2]{{%
\expandafter\ifx\csname l@#1\endcsname\relax
\typeout{** WARNING: IEEEtran.bst: No hyphenation pattern has been}%
\typeout{** loaded for the language `#1'. Using the pattern for}%
\typeout{** the default language instead.}%
\else
\language=\csname l@#1\endcsname
\fi
#2}}
\providecommand{\BIBdecl}{\relax}
\BIBdecl

\bibitem{Macoskey2021BifocalNA}
J.~J. Macoskey, G.~P. Strimel, and A.~Rastrow, ``Bifocal neural asr: Exploiting
  keyword spotting for inference optimization,'' \emph{ICASSP 2021 - 2021 IEEE
  International Conference on Acoustics, Speech and Signal Processing
  (ICASSP)}, pp. 5999--6003, 2021.

\bibitem{Panchapagesan2016MultiTaskLA}
S.~Panchapagesan, M.~Sun, A.~Khare, S.~Matsoukas, A.~Mandal, B.~Hoffmeister,
  and S.~Vitaladevuni, ``Multi-task learning and weighted cross-entropy for
  dnn-based keyword spotting,'' in \emph{INTERSPEECH}, 2016.

\bibitem{Compressedinproceedings}
M.~Sun, D.~Snyder, Y.~Gao, V.~Nagaraja, M.~Rodehorst, S.~Panchapagesan,
  N.~Strom, S.~Matsoukas, and S.~Vitaladevuni, ``Compressed time delay neural
  network for small-footprint keyword spotting,'' 08 2017, pp. 3607--3611.

\bibitem{Sun2017AnES}
M.~Sun, A.~Schwarz, M.~Wu, N.~Strom, S.~Matsoukas, and S.~Vitaladevuni, ``An
  empirical study of cross-lingual transfer learning techniques for
  small-footprint keyword spotting,'' \emph{2017 16th IEEE International
  Conference on Machine Learning and Applications (ICMLA)}, pp. 255--260, 2017.

\bibitem{Kumatani2017DirectMO}
K.~Kumatani, S.~Panchapagesan, M.~Wu, M.~Kim, N.~Strom, G.~Tiwari, and
  A.~Mandal, ``Direct modeling of raw audio with dnns for wake word
  detection,'' \emph{2017 IEEE Automatic Speech Recognition and Understanding
  Workshop (ASRU)}, pp. 252--257, 2017.

\bibitem{Bottleneckinproceedings}
J.~Guo, K.~Kumatani, M.~Sun, M.~Wu, A.~Raju, N.~Strom, and A.~Mandal,
  ``Time-delayed bottleneck highway networks using a dft feature for keyword
  spotting,'' 04 2018.

\bibitem{minhua1}
M.~{Wu}, S.~{Panchapagesan}, M.~{Sun}, J.~{Gu}, R.~{Thomas}, S.~N. {Prasad
  Vitaladevuni}, B.~{Hoffmeister}, and A.~{Mandal}, ``Monophone-based
  background modeling for two-stage on-device wake word detection,'' in
  \emph{2018 IEEE International Conference on Acoustics, Speech and Signal
  Processing (ICASSP)}, 2018, pp. 5494--5498.

\bibitem{Smallfootprint1}
G.~{Chen}, C.~{Parada}, and G.~{Heigold}, ``Small-footprint keyword spotting
  using deep neural networks,'' in \emph{2014 IEEE International Conference on
  Acoustics, Speech and Signal Processing (ICASSP)}, 2014, pp. 4087--4091.

\bibitem{sainath15}
T.~N. Sainath and C.~Parada, ``Convolutional neural networks for
  small-footprint keyword spotting.'' in \emph{INTERSPEECH}, 2015, pp.
  1478--1482.

\bibitem{omar_asru}
M.~O. Khursheed, C.~Jose, R.~Kumar, G.~Fu, B.~Kulis, and S.~K. Cheekatmalla,
  ``Tiny-crnn: Streaming wakeword detection in a low footprint setting,'' in
  \emph{2021 IEEE Automatic Speech Recognition and Understanding Workshop
  (ASRU)}, 2021, pp. 541--547.

\bibitem{Khursheed2020SmallFC}
M.~O. Khursheed, C.~Jose, R.~Kumar, G.~Fu, B.~Kulis, and S.~Cheekatmalla,
  ``Small footprint convolutional recurrent networks for streaming wakeword
  detection,'' \emph{arXiv: Audio and Speech Processing}, 2020.

\bibitem{Sun2016MaxpoolingLT}
M.~Sun, A.~Raju, G.~Tucker, S.~Panchapagesan, G.~Fu, A.~Mandal, S.~Matsoukas,
  N.~Strom, and S.~N.~P. Vitaladevuni, ``Max-pooling loss training of long
  short-term memory networks for small-footprint keyword spotting,'' \emph{2016
  IEEE Spoken Language Technology Workshop (SLT)}, pp. 474--480, 2016.

\bibitem{kws_1}
Y.~Wang, J.~Li, and Y.~Gong, ``Small-footprint high-performance deep neural
  network-based speech recognition using split-vq,'' in \emph{2015 IEEE
  International Conference on Acoustics, Speech and Signal Processing
  (ICASSP)}, 2015, pp. 4984--4988.

\bibitem{kws_2}
M.~Sun, V.~Nagaraja, B.~Hoffmeister, and S.~Vitaladevuni, ``Model shrinking for
  embedded keyword spotting,'' in \emph{2015 IEEE 14th International Conference
  on Machine Learning and Applications (ICMLA)}, 2015, pp. 369--374.

\bibitem{kws_3}
S.~Zhang, W.~Liu, and Y.~Qin, ``Wake-up-word spotting using end-to-end deep
  neural network system,'' in \emph{2016 23rd International Conference on
  Pattern Recognition (ICPR)}, 2016, pp. 2878--2883.

\bibitem{kws_4}
M.~Wu, S.~Panchapagesan, M.~Sun, J.~Gu, R.~Thomas, S.~N. Prasad~Vitaladevuni,
  B.~Hoffmeister, and A.~Mandal, ``Monophone-based background modeling for
  two-stage on-device wake word detection,'' in \emph{2018 IEEE International
  Conference on Acoustics, Speech and Signal Processing (ICASSP)}, 2018, pp.
  5494--5498.

\bibitem{Sigtia}
S.~Sigtia, J.~Bridle, H.~Richards, P.~Clark, E.~Marchi, and V.~Garg,
  ``Progressive voice trigger detection: Accuracy vs latency,'' in \emph{ICASSP
  2021 - 2021 IEEE International Conference on Acoustics, Speech and Signal
  Processing (ICASSP)}, 2021, pp. 6843--6847.

\bibitem{lcrnn}
H.~Du, R.~Li, D.~Kim, K.~Hirota, and Y.~Dai, ``Low-latency convolutional
  recurrent neural network for keyword spotting,'' in \emph{2018 Joint 10th
  International Conference on Soft Computing and Intelligent Systems (SCIS) and
  19th International Symposium on Advanced Intelligent Systems (ISIS)}, 2018,
  pp. 802--807.

\bibitem{Gengshen}
G.~Fu, T.~Senechal, A.~Challenner, and T.~Zhang, ``Unified speculation,
  detection, and verification keyword spotting,'' \emph{2022 IEEE International
  Conference on Acoustics, Speech and Signal Processing (ICASSP)}, 2022.

\bibitem{det}
\BIBentryALTinterwordspacing
S.~Z. Li and A.~Jain, Eds., \emph{DET Curves}.\hskip 1em plus 0.5em minus
  0.4em\relax Boston, MA: Springer US, 2009, pp. 223--223. [Online]. Available:
  \url{https://doi.org/10.1007/978-0-387-73003-5643}
\BIBentrySTDinterwordspacing

\bibitem{jose20_interspeech}
C.~Jose, Y.~Mishchenko, T.~Sénéchal, A.~Shah, A.~Escott, and S.~N.~P.
  Vitaladevuni, ``{Accurate Detection of Wake Word Start and End Using a
  CNN},'' in \emph{Proc. Interspeech 2020}, 2020, pp. 3346--3350.

\end{thebibliography}
